\documentclass{article}
\usepackage{epsfig}
\usepackage{amsmath}

\newcommand{\degree}{^\circ}
\newcommand{\onefigure}[2]{\begin{figure}[htbp]
\begin{center}\leavevmode\epsfbox{#1.eps}\end{center}\caption{#2\label{#1}}
\end{figure}}
\newcommand{\setfigure}[2]{\begin{figure}[htbp]
\begin{center}\leavevmode\epsfxsize=4in\epsfbox{#1.eps}\end{center}\caption{#2\label{#1}}
\end{figure}}

\newcommand{\be}{\begin{equation}}
\newcommand{\ee}{\end{equation}}

\begin{document}
\pagestyle{empty}

\bigskip\bigskip
\begin{center}
{\bf \large A Conjecture for Using
Optical Methods for Affecting Superfluid Hydrodynamics}
\end{center}

\begin{center}
James Lindesay\footnote{e-mail address, jlslac@slac.stanford.edu} \\
Computational Physics Laboratory \\
Howard University,
Washington, D.C. 20059 
\end{center}

\begin{center}
{\bf Abstract}
\end{center}
The relation between the macroscopic 
quantum coherent nature of superfluids and the coherent properties 
of optical interference patterns will be utilized to examine the
optical properties of superfluid hydrodynamics. A Bragg pattern 
imposed on the superfluid (either holographically 
or using a phase mask)  
is expected to induce periodic variations in 
the local index of refraction of the normal and super fluid components. 
The altered optical properties can then be probed by
a second coherent light source.  In this
manner, the behavior of the probe beam can be
switched using the specific characteristics of the imposed pattern. 
Acoustic modes should also manifest measurable affects on
incident coherent radiations.
\newpage

\setcounter{equation}{0}
\section{Introduction}
\indent

There is considerable interest in developing optical phase
masks with diffactive orders that can be tuned for varying
applications without a need for material modifications or
reconstruction.  Such devices would not only be useful
for writing Bragg patterns in optical fibers for communication
purposes, but if the pattern is optically induced, would
provided a mechanism for the optical switching of
propagated modes.  The development of such a device
is explored in what follows.  A more detailed
description can be found in the literature\footnote{{\small  Lindesay, James V.,
Lyons, Donald R., and Quiett, Carramah J. ``The Design of Fiber Optic
Sensors for Measuring Hydrodynamic Parameters'', \textit{Trends in
Electro-Optics Research}, William T. Arkin, Ed., Nova Science Publishers,
New York (ISBN 1-59454-498-0) (2006)}
\label{JLDLCQ}}.

The imposed pattern induces spatial dependence of 
the dielectric constant of the form 
\be
\epsilon =n_{o}^{2} +\delta \epsilon \,\cos ^{2} \left( \frac{\pi y}{w}
\right) ,
\label{dielectric}
\ee
where $\delta \epsilon$ is dependent upon the field 
intensity of the pattern through induced temperature and density 
perturbations.
The optical response of this pattern is then 
probed by directing a second coherent beam into this 
region, causing the probe beam to decompose into diffraction 
orders as illustrated in Figure \ref{probing}:
\begin{figure}[htbp]
\begin{center}
\includegraphics[width=0.547in, height=1.354in]{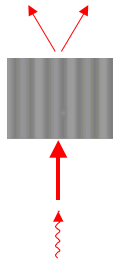}
\caption{Probing Optical Density Pattern in Superfluid}
\end{center}
\label{probing}
\end{figure}
As shown elsewhere, this response is effectively that
of a tunable phase mask\footnote{{\small Lyons, 
Donald R and Lindesay, James V.. ``Quantum Optical Mehtods of 
and Apparatuses for Writing Bragg Reflection Filters'', \textit{U.S. 
Patent 6,434,298} (Aug 2002)}}, with the probe beam optically 
switched by the holographic pattern.  The effect should
manifest for any non-linear optical material or macroscopic
quantum system, such as a superfluid or superconductor.  In
particular, the affects of such induced patterns on an
optically thin superfluid will be examined in what follows.

\setcounter{equation}{0}
\section{Electrodynamic Equations}
\indent

Maxwell's equations in a spatially varying
dielectric medium can be expressed in the form
\be
\begin{array}{l}
\vec{\nabla } (\vec{E} \cdot \vec{\nabla} log \, \epsilon ) +
\nabla ^2 \vec{E} - {\epsilon \over c^2} {\partial ^2 \vec{E}
\over \partial t^2} = 0, \\ \\
-\vec{\nabla } \, \epsilon \times \vec{E} +
\nabla^2 \vec{B} - {\epsilon \over c^2} {\partial ^2 \vec{B}
\over \partial t^2} = 0 .
\end{array}
\ee
The time derivatives of the log of the 
dielectric constant are assumed small compared to the frequency of the 
electromagnetic fields. 
The polarization $E_{x}$ of the probe field is 
chosen perpendicular to the induced variations in the dielectric $\epsilon(y)$. 

The probe field then reflects this periodic behavior
\be
E_{x} (y,z)\,=\, \sum\limits_{m}G_{m} (z)\,e^{i\frac{2m\pi }{w} y} 
\ee
Assuming a z-dependence of the form
\be
G_{m} (z;q)\,=\,A_{m} e^{iqz} \,+\,F_{m} e^{-iqz} 
\ee
results in an eigenvalue equation for the propagation constants $(q^2)$, 
with degenerate eigenvectors $A_m$ and $F_m$. Substitution of the
form in Eq. \ref{dielectric} gives the equation satisfied by the coefficients:
\be
\left\{ \epsilon _{0} \left( \frac{\omega }{c} \right) ^{2} -\left(
\frac{2\pi m}{w} \right) ^{2} -q^{2} \right\} A_{m} \,+\,\frac{\delta
\epsilon }{4} \left\{ A_{m+1} +A_{m-1} \right\} \,=\,0.
\ee
The unconstrained 
coefficients are chosen to satisfy the incoming and outgoing 
boundary conditions at the interphases.

To get a feel for the scale of the mode mixing in the probe
beam, consider a pattern with spacing $=6 \mu m$ imposed on 
a material with refractive index 1.75 inducing relative index variations 
of 0.5\%. If a probe beam of wavelength $0.632 \mu m$ is incident 
at the $+1^{st}$ order angle $6.04637\degree$, the resulting orders 
can be numerically solved as a function of sample thickness.
The amplitudes (assuming minimal beam absorption) are 
demonstrated in Figure \ref{Fig2}.
\onefigure{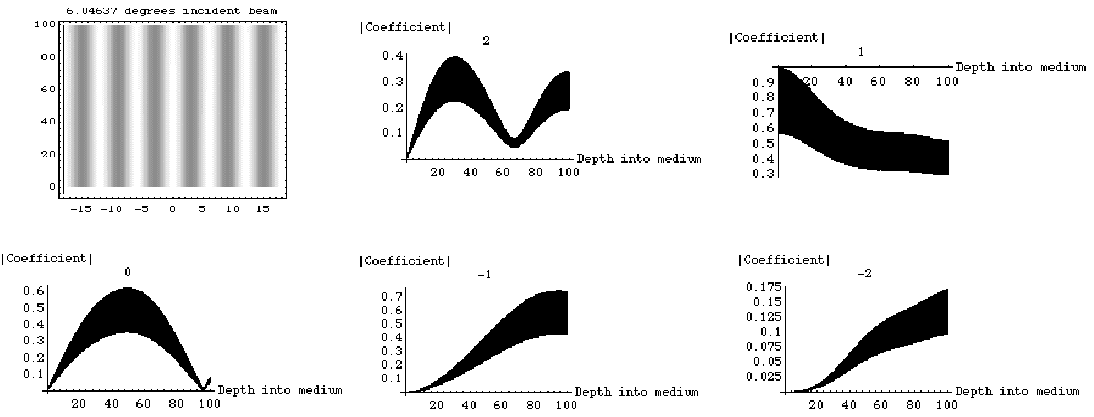}{Numerical form of probe order mixing due to 
periodic dielectric modification as a function 
of pattern depth $\mu m$}
For pattern depths greater than about 50 $\mu m$
there is significant mode mixing.

\setcounter{equation}{0}
\section{Optical Properties of Quantum Fluids}
\indent

Many of the hydrodynamic properties of liquid $^{4}$He can be understood 
in terms of a quantum two fluid model. The normal 
fluid has viscous flow and carries any entropy flux through 
the fluid, while below $T_\lambda \cong 2.17 \degree K$
there is a superfluid component that behaves like 
a macroscopic quantum system exhibiting persistent non-viscous 
flow and quantization of circulation\footnote{{\small D.R. Tilley and J. Tilley, \textit{Superfluidity 
and Superconductivity}, Adam Hilger, LTD, Bristol and Boston, 
2$^{nd}$ edition (1986)}}.
Helium forms a low density liquid ($\rho\sim0.15g/l$) with
the refractive density smoothly modeled using the 
form\footnote{\textit{{\small CRC Handbook of Physics and Chemistry}, David 
R. Lide (Editor in Chief), New York: CRC Press, 84$^{th}$ edition, 
2003-2004}}
\be
n_{He} \,\cong \,1.000\,+\,0.193 \,\rho /g\cdot cm^{-3} .
\label{index}
\ee

\subsection{Two fluid hydrodynamics}
\indent

The hydrodynamic flows of this system are described by
two fluid equations\footnote{{\small J. Lindesay and H. Morrison, ``The Geometry 
of Quantum Flow''\textit{,} in \textit{Mathematical Analysis of Physical 
Systems}, pp 135-167, edited. by R. Mickens. Van Nostrand Reinhold, 
Co., New York (1985)}}, which include the mass continuity equation
\be
\frac{\partial \rho }{\partial t} \,+\,\vec{\nabla } \cdot \left (
\rho_s \vec{v}_s + \rho_n \vec{v}_n \right ) ,
\label{masscontinuity}
\ee
the entropy flux equation
\be
\frac{\partial \left( \rho \sigma \right) }{\partial t} +\vec{\nabla } \cdot \left (
 \rho \sigma \vec{v}_n \right )  ,
\label{entropyflux}
\ee
the superfluid Euler equation
\be
\rho \left( \frac{\partial \vec{v}_s }{ \partial t } +
(\vec{v}_s \cdot \vec{\nabla} ) \, \vec{v}_s \right) =
-\vec{\nabla} P + \rho \, \sigma \, \vec{\nabla} T +
{\rho_n \over 2} \vec{\nabla} (\vec{v}_n - \vec{v}_s)^2 ,
\label{Euler}
\ee
and the Navier-Stokes equation
\be
\rho_s \left( \frac{\partial \vec{v}_s }{ \partial t } +
(\vec{v}_s \cdot \vec{\nabla} ) \, \vec{v}_s \right) +
\rho_n \left( \frac{\partial \vec{v}_n }{ \partial t } +
(\vec{v}_n \cdot \vec{\nabla} ) \, \vec{v}_n \right) =
-\vec{\nabla} P + \eta \nabla^2 \vec{v}_n +
{\eta \over 3} \vec{\nabla}(\vec{\nabla} \cdot \vec{v}_n) .
\label{NavierStokes}
\ee

Some relevant parameters of superfluid helium will
be given for future reference.  
The speed of (first) sound (ie. compressional waves) is given by 
$u_{1}^{2} \equiv \left( \frac{\partial P}{\partial \rho } \right)
_{\sigma }$, with typical values of the order $u_{1} \sim 220-240 \, m/s$.
Temperature waves in superfluids propagate with the
velocity of \textit{second sound} given by
$(u_{2})^{2} \,=\,\frac{\rho _{s} }{\rho _{n} } \frac{T\sigma ^{2} }{c_{P} }$.
Typical values for this speed are $u_2 \sim 20 \, m/s$. 
The entropy per unit mass is about
$\sigma \cong 100 \, m^{2} /s^{2} \, \degree K$.

\subsection{Standing waves in a bulk quantum fluid}
\indent

The strategy of the present approach is to utilize the
coherent nature of radiation from a laser to affect the
coherent behavior of a macroscopic quantum system.
Figure \ref{Fig3} represents an arrangement of a macroscopic
loop which establishes a set 
of discrete valued properties within the quantum fluid.
\onefigure{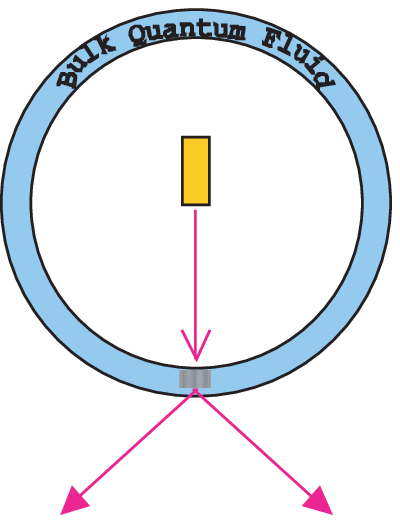}{Superimposed coherent quantum states}
The region being probed could be excited by mechanical, thermal,
or optical perturbations\footnote{{\small Lyons, Donald R. ``Apparatus 
for and methods of sensing evanescent events in a fluid field'', \textit{U.S. 
Patent 6,650,799} (Nov 2003) and \textit{U.S. Patent 6,915,028} (Jul 2005)}}. 
Generally, the superfluid component is locally accelerated from regions 
of high chemical potential towards regions of lower chemical 
potential.

Assume that the speeds $v_{s}$, $v_{n}$, the
density and entropy perturbations $\delta \rho, \delta \sigma$, 
and the pressure and temperature gradients are all first order small 
variations from equilibrium values. Stationary temperature variations
will be imposed on the quantum fluid 
in the form of the real part of
\be
T(x,t)\,=\,T_{o} +\frac{\delta T}{2} \left( 1+\cos \frac{2\pi x}{W}
\right) e^{-i\omega t} 
\ee
The perturbative forms of
Equations \ref{masscontinuity}-\ref{NavierStokes}
are then given by
\be
-\omega \, \delta \rho \,+\,\frac{2\pi }{W} \left( \rho _{s} v_{s} +\rho
_{n} v_{n} \right) \,=\,0 ,
\ee
\be
-\omega \, \left( \sigma \delta \rho +\rho \delta \sigma \right)
\,+\,\frac{2\pi }{W} \left( \rho \sigma v_{n} \right) \,=\,0 ,
\ee
\be
-\omega \, \rho v_{s} \,=\,-\frac{2\pi }{W} \delta P+\rho \sigma \frac{2\pi
}{W} \frac{\delta T}{2} +\rho _{n} \frac{2\pi }{W} \left( v_{n} -v_{s}
\right) ^{2} ,
\ee
\be
\omega \, \left( \rho _{s} v_{s} +\rho _{n} v_{n} \right) \,=\,\frac{2\pi
}{W} \delta P-i\eta \left( \left( \frac{2\pi }{W} \right) ^{2} v_{n}
+\frac{1}{3} \omega \left( \frac{2\pi }{W} \right) \frac{\sigma \delta
\rho +\rho \delta \sigma }{\rho \sigma } \right) .
\ee
The static limit \ensuremath{\omega}=0 gives the condition
$\delta P=\frac{1}{2} \rho \sigma \,\delta T$
known as the \emph{fountain effect}. Using the thermodynamic 
Maxwell relation
$\left( \frac{\partial P}{\partial \sigma } \right) _{\rho } =\rho ^{2}
\left( \frac{\partial T}{\partial \rho } \right) _{\sigma } =-\frac{\rho
}{\beta _{\sigma } } $,
the density perturbations can be expressed in terms of the speed 
of first sound and adiabatic thermal expansion coefficient
using
$\delta \rho \,=\,\frac{\sigma \left( \frac{1}{2} \rho \delta
T\,+\,\frac{\rho }{\beta _{\sigma } } \frac{\delta \sigma }{\sigma }
\right) }{u_{1}^{2} } $.
The density variations are therefore expected to satisfy
\be
\begin{array}{l l}
\delta \rho \thinspace  & = \thinspace \left( \frac{2\pi }{W} \right)
^{4} \frac{\sigma ^{2} }{\omega ^{2} } \left( \frac{\partial P}{\partial
\sigma } \right) _{\rho } \left( \frac{\rho _{s} }{\rho _{n} } \right)
\frac{\delta T}{2\left( \left( \frac{2\pi \thinspace u_{1} }{W} \right)
^{2} -\omega ^{2} \right) }  \\
 & = -\thinspace \left( \frac{2\pi }{W} \right) ^{4} \frac{\sigma ^{2}
}{\omega ^{2} } \left( \frac{\rho }{\beta _{\sigma } } \right) \left(
\frac{\rho _{s} }{\rho _{n} } \right) \frac{\delta T}{2\left( \left(
\frac{2\pi \thinspace u_{1} }{W} \right) ^{2} -\omega ^{2} \right) } .
\end{array}
\ee

The scale of the variations are seen to be significant for low 
frequencies and a small coefficient of volumetric expansion. 
The isobaric expansion coefficient for superfluid helium changes 
significantly, as plotted in Figure \ref{Fig4}.
\setfigure{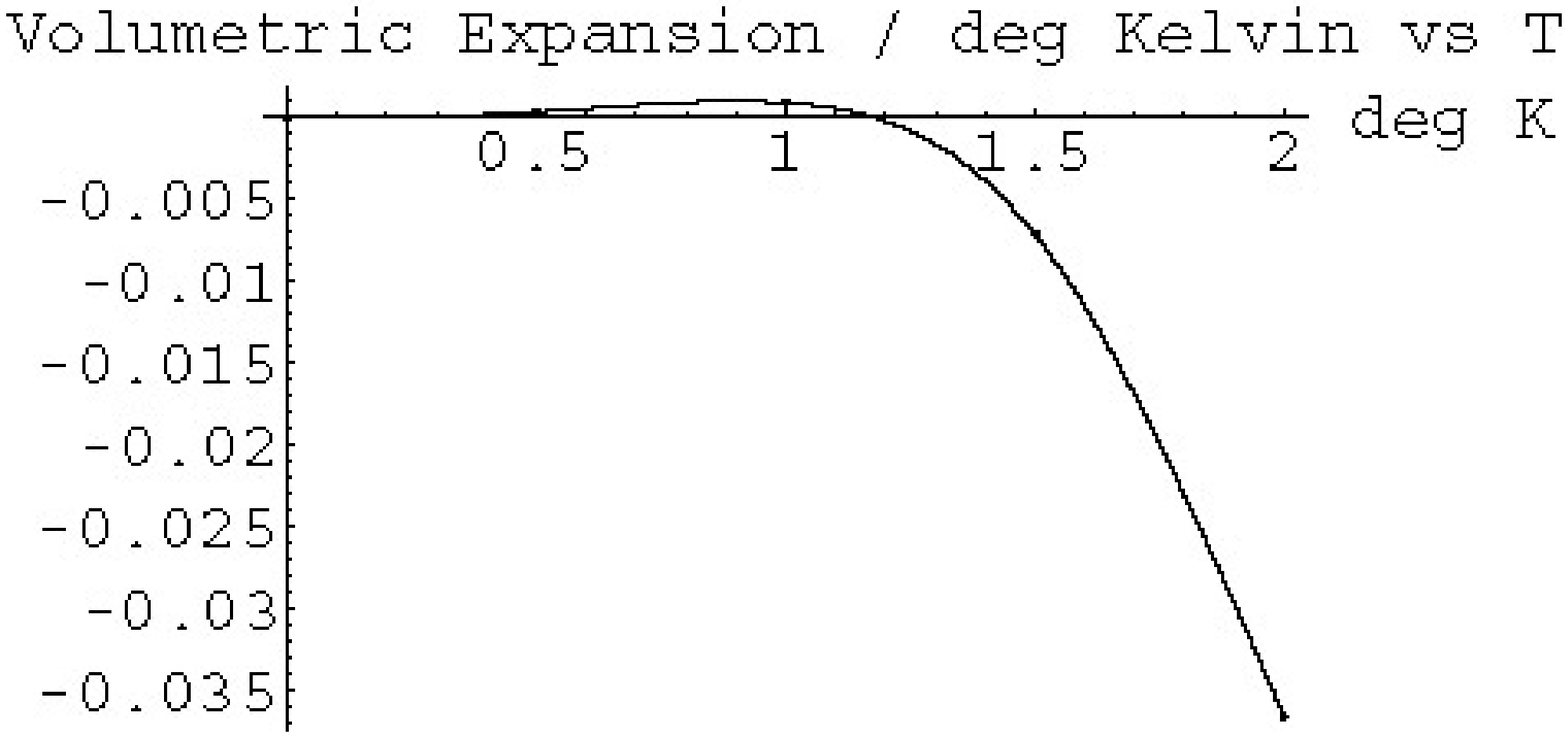}{Coefficient of Isobaric Volumetric Expansion for Liquid 
Helium}
For temperatures $T \sim1.5 ^\circ K$ where
$\beta \sim0.01/^\circ K$, and a typical optical pattern spacing of
$W \cong 10^{-6} m$, the density perturbations
for frequencies $f=2 \pi \omega$ are of the order 
\be
\frac{\delta \rho }{\rho } \,\sim \, \left \{
\begin{array}{l l}
\quad -70 \quad \frac{\delta T}{\degree K}
\left ( \frac{MHz }{ f } \right ) ^2 &
f << {u_1 \over W} \\ \\
4 \times 10^6 \, {\delta T \over \degree K} \left ( MHz \over f \right )^4 &
f >> {u_1 \over W}
\end{array} .
\right .
\ee
This effect is considerably enhanced for slightly lower temperatures
where the volumetric expansion coefficient becomes vanishingly
small.  Equation \ref{index} then relates the dimensionless 
relative index variation to liquid helium density variations,
which for low frequencies is of the order 
\be
\frac{\delta n}{n} \cong 0.029\frac{\delta \rho }{\rho } .
\ee
Since there is considerable
variability with temperature of the normal fluid 
density (which decreases to zero at absolute zero) and the coefficient 
of volumetric expansion, one expects to be able to arrange
conditions such that there are measurable effects upon 
the optical properties a superfluid due to an imposed
time varying pattern.

Coherent light should also modify the material 
properties of the quantum fluid, changing its mechanical and 
thermal states. In particular, we expect quantized resonant 
responses of superfluid systems to coherent perturbative influences 
when configured as in Figure \ref{Fig3}. 
Since as previously mentioned the speed of second sound is of order
$u_2 \sim 20 \, m/s$, there would be micron scale temperature
variations for frequencies of the order 200KHz, 
with lower frequencies requiring patterns of wider spacing. There should be resonant thermal 
wave effects for such patterns.  Standing acoustic waves of
micron scale wavelengths should immediately have
measurable effects on the probe beam, assuming the
wave pattern can be appropriately stabilized.

\setcounter{equation}{0}
\section{Conclusion}
\indent

The two fluid model suggests that an optical interference pattern 
placed on a superfluid should induce local variations 
in the fluid's hydrodynamic parameters. A second coherent
light source can be used to probe those variations through
diffractive effects.  Calculations have been presented that
suggest the scale of hydrodynamic variations that can be induced,
and the potential measurement of those variations.  Any
measured diffractive effect on a probe beam due to the
presence of an imposed pattern would demonstrate an
optical switching of that probe beam.  It is likewise
suggested that coherent mechanical perturbations
of appropriate scale in
the superfluid should have a
measurable impact on the properties of the probe beam.

\begin{center}
\textbf{Acknowledgements}
\end{center}

The author would like to acknowledge useful discussions with
Donald R. Lyons of the Fiber 
Optic Sensors and Smart Structures group in the Research Center 
for Optical Physics at Hampton University.

\end{document}